\documentclass[apjl]{emulateapj}
\usepackage{apjfonts}

\usepackage{amsmath}

\newcommand{\umjy}{$\mu$Jy}

\newcommand{\cpks}{ct~ks$^{-1}$}
\newcommand{\kms}{km~s$^{-1}$}

\newcommand{\ergps}{erg~s$^{-1}$}
\newcommand{\ergpshz}{erg~s$^{-1}$~Hz$^{-1}$}

\newcommand{\hr}{\mbox{$^{\mathrm h}$}}
\newcommand{\mn}{\mbox{$^{\mathrm m}$}}
\newcommand{\chan}{\textit{Chandra}}

\slugcomment{To appear in the Astrophysical Journal Letters}

\shorttitle{A Deep Look at $\epsilon$ Ind Bab}
\shortauthors{Audard et al.}

\begin{document}


\title{A Deep Look at the T-Type Brown Dwarf Binary $\epsilon$ Indi B\lowercase{ab} with \chan\ 
and ATCA}

%
%
%
%

\author{M.~Audard\altaffilmark{1}, A.~Brown\altaffilmark{2}, K.~R.~Briggs\altaffilmark{3}, M.~G\"udel\altaffilmark{3}, 
A.~Telleschi\altaffilmark{3}, and J.~E.~Gizis\altaffilmark{4}}

\altaffiltext{1}{Columbia Astrophysics Laboratory, Mail code 5247, 550 West
120$^\mathrm{th}$ Street, New York, NY 10027, {audard@astro.columbia.edu}}

\altaffiltext{2}{Center for Astrophysics and Space Astronomy, University of Colorado, Boulder, CO 80309-0389, {ab@casa.colorado.edu}}
\altaffiltext{3}{Paul Scherrer Institut, Villigen \& W\"urenlingen, 5232 Villigen PSI, Switzerland, {briggs@astro.phys.ethz.ch, 
guedel@astro.phys.ethz.ch, atellesc@astro.phys.ethz.ch}}
\altaffiltext{4}{Department of Physics and Astronomy, 223 Sharp Lab, University of Delaware, Newark, DE 19716, {gizis@udel.edu}}

\begin{abstract}
We present deep observations of the nearby T-type brown dwarf binary \objectname{$\epsilon$ Indi Bab} in radio with the Australia Telescope
Compact Array and in X-rays with the \chan\  X-ray Observatory. Despite long integration times, the binary (composed of T1 and T6 dwarfs) 
was not detected in either wavelength regime. We reached $3\sigma$ upper limits of $1.23 \times 10^{12}$ and $1.74 \times 10^{12}$~\ergpshz\ 
for the radio luminosity at 4.8~GHz and 8.64~GHz, respectively; in the X-rays, the upper limit in the $0.1-10$~keV band
was $3.16 \times 10^{23}$~\ergps. We discuss the above results in the framework of magnetic activity in 
ultracool, low-mass dwarfs.

\end{abstract}

\keywords{radio continuum: stars---stars: activity---stars: coronae---stars: individual ($\epsilon$ Ind Bab)---stars: low-mass, brown dwarfs---X-rays: stars}

\section{Introduction}
\label{sect:intro}

Magnetic activity is a common feature of late-type stars. The general
hypothesis assumes that some magnetic dynamo mechanism ($\alpha\omega$ dynamo)
occurs at the interface between the radiative core of the star and its outer convective zone.
However, objects with masses below 0.3~$M_\odot$ (spectral type later than M3) are generally believed to be 
fully convective, and thus cannot support the $\alpha\omega$ dynamo.
Nevertheless, magnetic activity is commonly seen, even in very late M-type
objects \citep[e.g.,][]{johnson81,johnson87,tagliaferri90,drake96,fleming00,rutledge00,fleming93,fleming03,stelzer04}.
An alternative dynamo mechanism ($\alpha^2$ dynamo),
involving turbulent magnetic fields, has been proposed to take place in 
fully convective objects \citep[e.g.,][]{durney93}.

Studies of the H$\alpha$ emission (a common tracer of chromospheric activity) in late-type stars  have shown
a steep decrease for spectral types later than M7 despite their fast rotation \citep{gizis00,mohanty03}.
The X-ray emission also appears to decrease, although the sample of studied stars is limited
\citep[e.g.,][]{fleming93,fleming03}. It was thus a surprise when \citet{berger02} found no decline in radio 
activity between spectral types M3 and L3.5, and even suggested a possible increase in radio luminosity with cooler 
effective temperature. Recent results seem to support such a trend in late-M/early-L dwarfs 
\citep{berger05,burgasser05}. Old T-type brown dwarfs are, therefore, attractive objects to observe, in order to
understand magnetic activity in ultracool dwarfs. However, observations of T dwarfs are scarce in the radio and 
were nonexistent in X-rays.

This paper extends the previous investigations into the T dwarf domain. We present the
first observation of a T dwarf in X-rays with the \chan\  X-ray Observatory along with a deep
observation in the radio regime with the Australia Telescope Compact Array (ATCA). Our target, $\epsilon$ Ind Bab, is the closest known brown dwarf 
and is composed of two objects with spectral types T1 and T6 separated by $0\farcs 73$ ($2.65$~AU at a distance of
$3.626$~pc; \citealt{scholz03,smith03,volk03,mccaughrean04}). We list the binary's main characteristics in
Table~\ref{tab:epsind}. \citet{blank05} failed to detect $\epsilon$ Ind Bab with ATCA, but our
radio observation goes deeper.

\section{Observations and Data Reduction}
\label{sect:data}

Despite our best effort to coordinate the \chan\  and ATCA observations of the $\epsilon$ Ind Bab binary, 
the X-ray observations were delayed by a few days due to satellite safety reasons. A log of the observations
is given in Table~\ref{tab:log}.

\subsection{ATCA}
The ATCA was in a long-baseline configuration (6D); we used 4.8~GHz and 8.64~GHz receivers with
bandwidths of 128~MHz. Observing scans ranged from 10~min to 20~min on source,
depending on the weather conditions, whereas we used 3~min scans for the phase calibrator.
About 5 minutes at the start of each observing round were spent on the flux calibrator.
The observing conditions on the first day were average with cloud coverage; however, the last three hours
of the first day of observation were essentially useless due to strong winds and a thunderstorm.
In contrast, the weather conditions were much better during the second day with generally
a cloud-free sky.

\begin{figure}
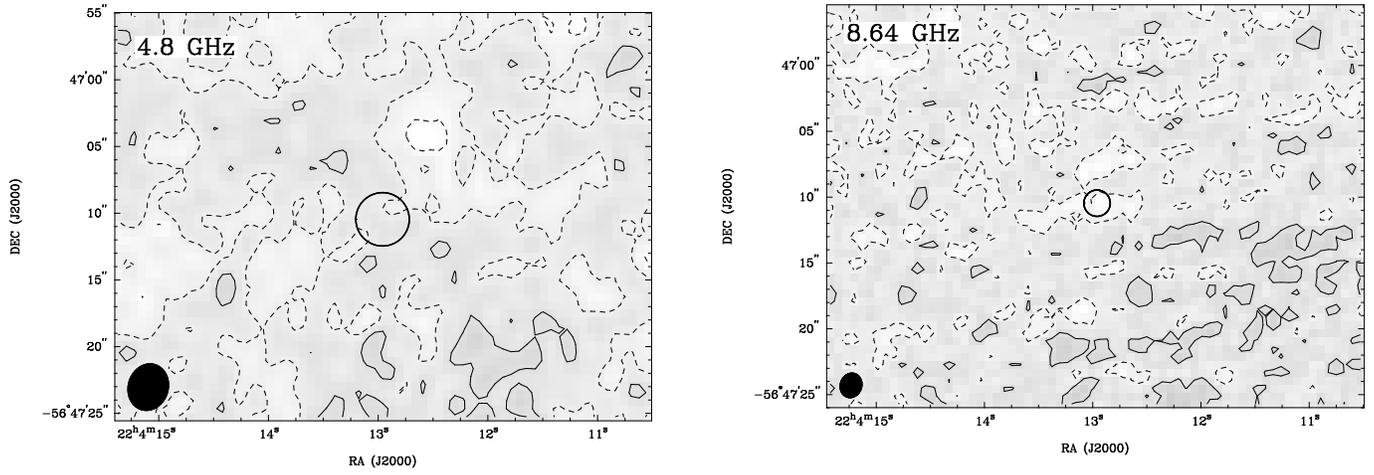

\centering
\includegraphics[angle=-90,width=0.475\textwidth]{f1a.eps}
\hfill
\includegraphics[angle=-90,width=0.475\textwidth]{f1b.eps}
\caption{Cleaned ATCA maps of the $\epsilon$ Ind Bab field at 4.8~GHz (left) and 8.64~GHz (right). Contours
are plotted in units of $-3,-1,1,3,10$ times $26.4$~\umjy\  and $37.3$~\umjy, respectively. The expected
position of the brown dwarf binary (Tab.~\ref{tab:log}) is shown by open circles of 2\arcsec\  and 1\arcsec\  radii, respectively. 
The beam size is also shown in the bottom left part of each map.
\label{fig:atcamaps}}
\end{figure}

We combined both observing rounds and reduced the ATCA data using the MIRIAD software \citep{sault95}. 
We detected 9 sources in the 4.8~GHz map; we used boxes of about 20\arcsec\  width centered on these sources
and applied a CLEAN algorithm using uniform weighting. About $5.4$~mJy were thus removed
in 202 iterations (we stopped when a negative value was encountered). Although no sources
were visible in the dirty map at 8.64~GHz, we used the boxes around the first five brightest sources
detected at longer wavelengths and performed a CLEAN algorithm.  About $0.5$~mJy were removed
after 38 iterations. The rms noise level in the cleaned maps reached $26.4$ and $37.3$~\umjy\ 
at 4.8 and 8.64~GHz, respectively. The values are close to the theoretical values 
($25.1$ and $35.6$~\umjy). No source was detected at the expected position of the T-type brown dwarf
binary (see Tab.~\ref{tab:log} and Fig.~\ref{fig:atcamaps}). Since short duration flares could be missed in the 2-day images,
we have generated maps down to 1-hour duration
but still detected no source (typical 
rms noise levels of $120$~\umjy\  and $155$~\umjy\  at 4.8 and 8.64~GHz, respectively).
Our non-detection limits are consistent with those of \citet{blank05} but we went deeper since we observed $\epsilon$
Ind Bab about twice as long. At the distance of the binary ($3.626$~pc), the $3\sigma$ upper limits
correspond to radio luminosities, $L_\mathrm{R}$, of $1.23 \times 10^{12}$ and $1.74 \times 10^{12}$~\ergpshz\  
at 4.8 and 8.64~GHz, respectively.

\subsection{\chan}
\begin{figure}
\centering
\includegraphics[angle=0,width=0.45\textwidth]{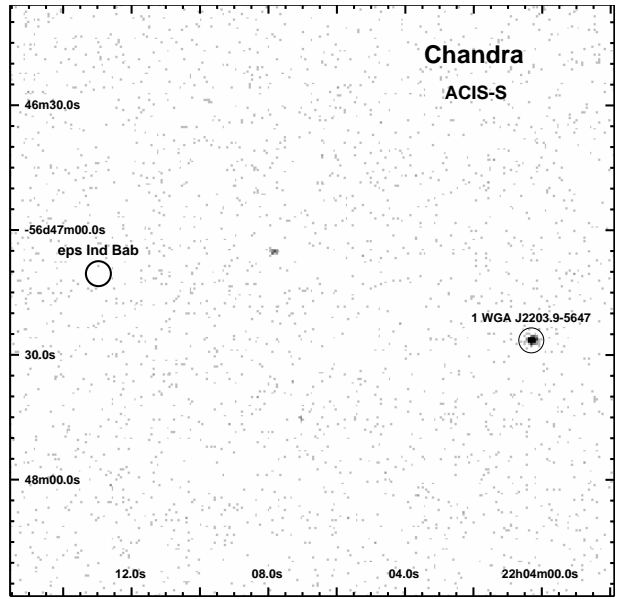}
\caption{Extract of the \chan\  ACIS-S image. The expected position of $\epsilon$ Ind Bab (Tab.~\ref{tab:log}) is shown with a circle of
3\arcsec\  radius.\label{fig:chandraim}}
\end{figure}

\chan\  observed $\epsilon$ Ind Bab at two different times. We reduced both data sets with the CIAO~3.1 software in
combination with CALDB 2.29. The data were taken in VERY FAINT mode which allowed us to
further reduce the background level. We later merged both event files into one single event file (total exposure
of $62.5$~ksec). No signal
was detected at the expected position of $\epsilon$ Ind Bab (Tab.~\ref{tab:log}); this was confirmed with several source detection
algorithms. Figure~\ref{fig:chandraim} shows an extract of the \chan\  ACIS-S image ($0.2-8$~keV).
\citet{blank05} tentatively associated the \textit{ROSAT} source 1WGA J220-5647 with $\epsilon$ Ind Bab.
However, thanks the high spatial resolution of \chan, we assign the \textit{ROSAT} source to a nearby 
bright X-ray source instead (see Fig.~\ref{fig:chandraim}). No optical or near-infrared counterpart
to this source could be found in Digitized Sky Survey and Two Micron All Sky Survey images.

No events were detected within $1\farcs 4$  (95\% of the encircled energy at 0.3~keV) 
of the expected position of $\epsilon$ Ind Bab in the $0.2-8$~keV range. Note that, without the energy
filtering, 4 events were detected but all had energies above 10~keV casting severe doubt that
they were due to the source rather than the background. Furthermore, increasing the extraction radius to 3\arcsec\  
only allows the detection of a single event. To estimate the background contribution,
we used a concentric annulus with an inner radius of $3$\arcsec\  and an outer radius calculated such 
that the annulus' area was 100 times larger than the area of the circle around the binary 
(i.e, $r_\mathrm{out} = 14\farcs 3$). A total of 77 events were detected; consequently, the scaled estimated background
contribution is $0.77$ events, i.e., a mean background rate of $0.012$~\cpks. We then followed
the approach of \citet{kraft91} to determine the upper confidence limit using
a Bayesian confidence level of 95\% which corresponded to $2.99$ counts. To convert this count rate of
$0.048$~\cpks\  into an X-ray luminosity, we used one plasma component with solar photospheric abundances
\citep{grevesse98} using the APEC 1.3.1 code \citep{smith01} in the XSPEC software \citep{arnaud96}. 
We obtained count rates in the $0.2-8$~keV
band and X-ray fluxes in the $0.1-10$~keV band. The latter fluxes were converted into X-ray luminosities, $L_\mathrm{X}$,
using the distance of $\epsilon$ Ind Bab. For plasma temperatures of $0.4$ to $1.0$~keV, the conversion
factor is constant, i.e., $6.5 \times 10^{24}$~\ergps\  per \cpks. For cooler plasma temperatures,
the conversion factor increases by factors of $1.16, 1.46, 1.74$, and $3.25$ for $kT = 0.3, 0.2, 0.15$, and $0.1$~keV,
respectively. Similarly, the factor increases by $1.20, 1.33, 1.53$, and $1.82$ for $kT = 1.5, 2, 3$, and $5$~keV, respectively.
Therefore, assuming $kT=0.4-1.0$~keV, we obtain $L_\mathrm{X} \la 3.16 \times 10^{23}$~\ergps. In the worst case ($kT=0.1$~keV),
our upper limit increases to $1.0 \times 10^{24}$~\ergps.

\section{Discussion}
\label{sect:disc}

Since the ATCA could not separate the binary and \chan\  would have barely done so, our upper limits ($L_\mathrm{R}$ and 
$L_\mathrm{X}$) can either be attributed to one binary component or the other. Table~\ref{tab:lum} gives the upper limits
of the luminosity ratios to the bolometric luminosity for each component.

Our radio upper limits, $L_\mathrm{R}/L_\mathrm{bol}<10^{-16.8}-10^{-16.0}$~Hz$^{-1}$, do not go as deep as for dMe stars, which are typically
detected with ratios of the order of $10^{-18}$~Hz$^{-1}$. However, the limit at 8.64~GHz is deeper for $\epsilon$ Ind Ba than the 
upper limit obtained at 8.46~GHz with the VLA by \citet{berger02} for a T6 brown dwarf (our limit at 8.64~GHz for $\epsilon$ 
Ind Bb is, however, equivalent). Berger hypothesized that the radio luminosity relative to the bolometric luminosity would increase
with later spectral type, and their strong detection of an L dwarf in radio
appears to support this \citep{berger05}. 
Similarly, despite the lack of sensitivity of current radio instruments, such a trend is also suggested in a small
survey of southern late-M and L dwarfs \citep{burgasser05}. On the other hand, 
our upper limits for T dwarfs do not indicate that this trend continues into the T dwarf
domain.
Deep observations of T dwarfs can, in principle, be very useful since they provide data points at low $T_\mathrm{eff}$;
however, the exposure required to achieve the necessary sensitivity for comparison with dMe dwarfs 
(i.e., $L_\mathrm{R}/L_\mathrm{bol}<10^{-18}-10^{-19}$~Hz$^{-1}$) must await the advent of the next generation of radio telescopes.

In contrast, the sensitivity of \chan\  allows us to obtain a low $L_\mathrm{X}/L_\mathrm{bol}$ ratio. 
We reach an upper limit about $10-100$ times lower than the $L_\mathrm{X}/L_\mathrm{bol}$
ratios observed in active dMe stars \citep[e.g.,][]{fleming95,fleming03}. In fact, our limit is close to the ratio
observed in the Sun (as a star) at its activity maximum and to the ratio observed in early-dM stars that do not show 
H$\alpha$ in emission \citep{fleming95}. 
Although X-ray observations of ultracool dwarfs are
rare \citep[e.g.,][]{rutledge00,schmitt02,martin02,fleming03,briggs04,stelzer04,berger05}, the X-ray emission in such dwarfs appears to
decline like the H$\alpha$ emission. Most X-ray detections were actually obtained during flares, with only a few
objects detected in quiescence. Our non-detection of the T-type $\epsilon$ Ind Bab with \chan\  reinforces the view that the X-ray
emission in ultracool dwarfs declines significantly with later spectral type.

\citet{berger05} suggested that magnetic activity is strong in ultracool dwarfs, in fact much stronger than in dMe
stars, but that it manifests itself mostly in the radio, as atmospheric conditions in such dwarfs become unfavorable 
for H$\alpha$ and X-ray emission due to the decoupling of the magnetic field from the neutral photospheric gas \citep{meyer99,mohanty02}. 
Taken alone, the lack of X-rays from $\epsilon$ Ind Bab could, in principle, support this view. However, \citet{berger02} noted that fast rotators 
($>10$~\kms) have high $L_\mathrm{R}/L_\mathrm{bol}$ ratios. But $\epsilon$ Ind Bab does not follow this
trend: \citet{smith03} found that the T1 component is a fast rotator ($v \sin i = 28$~\kms). 
Note that the stellar radius and period are similar to those 
of the L3.5 dwarf detected in radio (but not in X-rays or H$\alpha$) by \citet{berger05}.
Possibly, the suggested trend observed in late-M/early-L dwarfs does not hold in the old, cool T dwarfs ($T_\mathrm{eff} = 
800-1300$~K). Our observation thus suggests a physical change in magnetic behavior somewhere in the range of T dwarfs.
Alternatively, the observed sample of L and T dwarfs is too limited and sensitivity is lacking.

Magnetically active stars often show a relation between the quiescent X-ray and radio luminosities \citep{guedel93}.
The G\"udel-Benz relation points toward a close connection between non-thermal electrons observed as gyrosynchrotron
emission and the thermal plasma radiating in X-rays. The origin of this connection remains unclear; however,
there is increasing evidence that the X-ray emission in magnetically active stars is
due to the statistical superposition of flares \citep[e.g.,][]{audard00,guedel03}. In addition,
the radio and X-ray light curves of flares are often correlated \citep[e.g.,][]{neupert68,guedel02}. Thus, 
coronal heating by flares provides an elegant explanation for the time-average $L_\mathrm{X}-L_\mathrm{R}$ 
relation observed in active stars. However, ultracool dwarfs seem not to follow this relation
since they show stronger radio emission with respect to the X-ray emission \citep[e.g.,][]{berger01,berger02,burgasser05}. 
This indicates either that magnetic activity in such dwarfs is different \citep{berger01,berger05}, or possibly that the radio emission 
mechanism differs significantly from gyrosynchrotron. Indeed, the emission of
many flares in M dwarfs can be ascribed to 
a coherent mechanism \citep[e.g.,][]{kundu88}. In the case of $\epsilon$ Ind Bab, we obtained only upper limits for both radio and X-ray regimes; 
thus it is unclear whether the T brown dwarfs follow the G\"udel-Benz relation or not.

\section{Conclusions}
\label{sect:concl}

We have presented deep observations of $\epsilon$ Ind Bab, a T1+T6 brown dwarf binary at 3.626~pc, in X-rays with
\chan\  and in the radio with ATCA. The binary remained undetected, with upper limits of $L_\mathrm{X}/L_\mathrm{bol}
\la 10^{-5.4}$ and $10^{-4.7}$ and $L_\mathrm{R}/L_\mathrm{bol} \la 10^{-16.8}$ and $10^{-16.2}$~Hz$^{-1}$ (at 4.8~GHz)
for $\epsilon$ Ind Ba and Bb, respectively. The non-detection in the radio
is in contrast with the trend of increasing $L_\mathrm{R}/L_\mathrm{bol}$ with later spectral type suggested by \citet{berger02} and
\citet{berger05} in late-M and L dwarfs. \citet{berger05} argued that the neutral atmospheres in ultracool dwarfs lead to unfavorable conditions 
for the X-ray and H$\alpha$ emissions \citep{meyer99,mohanty02}, but not for the radio emission of fast rotators. However, the T1 component
in $\epsilon$ Ind Bab is a rapid rotator with similar period and radius as the L dwarf detected strongly in the
radio (but undetected in X-rays and H$\alpha$) by \citet{berger05}. Therefore, our $\epsilon$ Ind Bab observations 
indicate that cool T dwarfs may not follow the trend suggested by \citeauthor{berger02} and colleagues. However, a definitive
conclusion is difficult to reach in view of the limited sample of ultracool dwarfs observed in
the radio and in X-rays and of the lack of sensitivity of present radio (and X-ray) instruments.

\acknowledgments

The Australia Telescope Compact Array is part of the Australia Telescope which is funded by the
Commonwealth of Australia for operation as a National Facility managed by CSIRO.
Support for this work was provided by the National Aeronautics and Space Administration (NASA) through \chan\  Award 
Number G04-5002X to Columbia University and issued by the \chan\  X-ray Observatory Center, which is operated by the Smithsonian Astrophysical 
Observatory for and on behalf of the NASA under contract NAS8-03060, and by National Science Foundation grant AST-0206367 
and NASA grant NAG5-13058 to the University of Colorado. The PSI group acknowledges support from the 
Swiss National Science Foundation (grants 20-58827.99 and 20-66875.01).
We thank two referees (M.~Giampapa and an anonmyous one) for useful comments
and suggestions that improved the content of this paper. 
M.~A. thanks Scott Wolk and Bob Sault for their efforts to coordinate the \chan\
and ATCA observations.


\clearpage

\begin{deluxetable}{lcc}
\tabletypesize{\small}
\tablecolumns{3}
\tablewidth{0pc}
\tablecaption{The $\epsilon$ Ind B\lowercase{ab} binary\label{tab:epsind}}
\tablehead{
\colhead{\makebox[35mm][l]{}} & \colhead{$\epsilon$ Ind Ba} & \colhead{$\epsilon$ Ind Bb}}
\startdata
Spectral type\dotfill			& T1		& T6\\
$\log (L_\mathrm{bol}/L_\sun)$\dotfill& $-4.71$	& $-5.35$\\
$T_\mathrm{eff}$ (K)\dotfill		& $1276 \pm 35$	& $854 \pm 20$\\
Mass ($M_\mathrm{Jup}$)\dotfill		& $47 \pm 10$	& $28 \pm 7$\\
Radius ($R_\sun$)\dotfill		& $0.091 \pm 0.005$	& $0.096 \pm 0.005$\\
Distance (pc)\dotfill			& \multicolumn{2}{c}{$3.626$}\\
Age  (Gyr)\dotfill			& \multicolumn{2}{c}{$0.8-2$}
\enddata
\tablerefs{See \citet{mccaughrean04}}
\end{deluxetable}


\clearpage

\begin{deluxetable}{lc}
\tabletypesize{\footnotesize}
\tablecolumns{2}
\tablewidth{0pc}
\tablecaption{Observation Log for ATCA  and \chan\label{tab:log}}
\tablehead{
\multicolumn{2}{c}{Position of $\epsilon$ Ind Bab (Equinox: J2000; Epoch:  J2004.93)}}
\startdata
Right ascension 		&	$22\hr 04\mn 12\fs 96$\\
Declination			&	$-56\arcdeg 47\arcmin 10\farcs 5$\\
\cutinhead{ATCA (Program C1269)}
Observation date		&	2004 Dec 4 0h30--14h30 UT\\
{} 				&	2004 Dec 5 1h00--13h00 UT\\
Antenna configuration		&	6D\\
Bandwidth			& 	128 MHz over 32 channels\\
Flux calibrator			& 	PKS J1939$-$6342\\
\hspace*{2mm} (ATCA calibrator name)	&	(1934$-$638)\\
Phase calibrator		&	PMN J2121$-$6111\\
\hspace*{2mm} (ATCA calibrator name)	&	(2117$-$61)\\
{}				\\
8.64 GHz			\\
Synthesized beam		& $1\farcs 90 \times 1\farcs 65$\\
\quad Position angle		& $-18\fdg 7$\\
Theoretical RMS flux	 	&	35.56~\umjy\\
{}				\\
4.80 GHz			\\
Synthesized beam		& $3\farcs 55 \times 3\farcs 01$\\
\quad Position angle		& $-16\fdg 7$\\
Theoretical RMS flux		&	25.09~\umjy\\
\cutinhead{\chan\  (ObsId 6171 and 4485)}
Observation date		&	2004 Dec 7 13h59--21h11 UT\\
{}				&	2004 Dec 9 00h04--11h35 UT\\
Instrument			&	ACIS-S\\
Total exposure			&	62.4~ksec\\
PSF (FWHM)			&	$0\farcs 5$	\\
95\% EE PSF @ 0.3~keV		&	$1\farcs 44$	
\enddata
\end{deluxetable}


\clearpage

\begin{deluxetable}{lll}
\tabletypesize{\normalsize}
\tablecolumns{3}
\tablewidth{0pc}
\tablecaption{Upper limits of luminosities and of luminosity ratios to the bolometric luminosity\label{tab:lum}}
\tablehead{
\colhead{\makebox[4.0cm][c]{Upper limit}}  & \colhead{$\epsilon$ Ind Ba}   & \colhead{$\epsilon$ Ind Bb} }
\startdata
$L_{\mathrm{R,4.8~GHz}}$ (\ergpshz)\dotfill  				& \multicolumn{2}{c}{$1.23 \times 10^{12}$}\\
$L_{\mathrm{R,8.64~GHz}}$ (\ergpshz)\dotfill  				& \multicolumn{2}{c}{$1.74 \times 10^{12}$}\\
$L_{\mathrm{X,0.1-10~keV}}$ (\ergps)\dotfill    			& \multicolumn{2}{c}{$3.16 \times 10^{23}$}\\
$\log (L_{\mathrm{R,4.8~GHz}}/L_\mathrm{bol})$ (Hz$^{-1}$)\dotfill	& $-16.79$\tablenotemark{a}	&	$-16.15$\tablenotemark{a}\\
$\log (L_{\mathrm{R,8.64~GHz}}/L_\mathrm{bol})$ (Hz$^{-1}$)\dotfill	& $-16.64$\tablenotemark{a}	& 	$-16.00$\tablenotemark{a}\\
$\log (L_{\mathrm{X,0.1-10~keV}}/L_\mathrm{bol})$\dotfill		& $-5.38$\tablenotemark{a}	&	$-4.74$\tablenotemark{a}
\enddata
\tablenotetext{a}{Ratios calculated by attributing the luminosity upper limits to each component.}
\end{deluxetable}


\clearpage

\clearpage

\begin{thebibliography}{}

\bibitem[Arnaud(1996)]{arnaud96} Arnaud, K.~A. 1996, in ASP Conf. Ser. 101,
 Astronomical Data Analysis Software and Systems V, ed. G. Jacoby \& J. Barnes
 (San Francisco: ASP), 1

\bibitem[Audard et al.(2000)]{audard00} Audard, M., G{\" u}del, M., Drake, J.~J., \& Kashyap, 
V.~L.\ 2000, \apj, 541, 396 

\bibitem[Berger(2002)]{berger02} Berger, E.\ 2002, \apj, 572, 503 

\bibitem[Berger et al.(2001)]{berger01} Berger, E., et al.\ 2001, \nat, 410, 338

\bibitem[Berger et al.(2005)]{berger05} Berger, E., et al.\ 2005, \apj, submitted, astro-ph/0502384

\bibitem[Blank(2005)]{blank05} Blank, D.~L. 2005, \mnras, 354, 913

\bibitem[Briggs \& Pye(2004)]{briggs04} Briggs, K.~R., \& Pye, J.~P.\ 2004, \mnras, 353, 673 

\bibitem[Burgasser \& Putnam(2005)]{burgasser05} Burgasser, A.~J., \& Putnam, M.~E.\  2005, \apj, in press, astro-ph/0502365

\bibitem[Drake et al.(1996)]{drake96} Drake, J.~J., Stern, R.~A., Stringfellow, G., Mathioudakis, M., 
Laming, J.~M., \& Lambert, D.~L.\ 1996, \apj, 469, 828

\bibitem[Durney et al.(1993)]{durney93} Durney, B.R., De Young, B.S., \& Roxburgh, I.W. 1993, \solphys, 145, 207

\bibitem[Fleming et al.(2003)]{fleming03} Fleming, T.~A., Giampapa, M.~S., \& Garza, D. 2003, \apj, 594, 982

\bibitem[Fleming et al.(2000)]{fleming00} Fleming, T.~A., Giampapa, M.~S., \& Schmitt, J.~H.~M.~M.\ 2000, \apj, 533, 372

\bibitem[Fleming et al.(1993)]{fleming93} Fleming, T.~A., Giampapa, M.~S., Schmitt, J.~H.~M.~M., 
\& Bookbinder, J.~A. 1993,  ApJ, 410, 387

\bibitem[Fleming et al.(1995)]{fleming95} Fleming, T.~A., Schmitt, J.~H.~M.~M., \& Giampapa, M.~S.\ 1995, 
\apj, 450, 401 

\bibitem[Gizis et al.(2000)]{gizis00} Gizis, J.~E., Monet, D.~G., Reid, I.~N., 
Kirkpatrick, J.~D., Liebert, J., \& Williams, R.~J.\ 2000, \aj, 120, 1085 

\bibitem[Grevesse \& Sauval(1998)]{grevesse98} Grevesse, N., \& Sauval, A.~J.
1998, Space Science Reviews, 85, 161

\bibitem[G\"udel \& Benz(1993)]{guedel93} G\"udel, M., \& Benz, A.~O.\  1993, \apj, 405, L63

\bibitem[G{\" u}del et al.(2003)]{guedel03} G{\" u}del, M., Audard, M., Kashyap, V.~L., Drake, J.~J., \& 
Guinan, E.~F.\ 2003, \apj, 582, 423

\bibitem[G{\" u}del et al.(2002)]{guedel02} G{\" u}del, M., Audard, M., Skinner, S.~L., \& Horvath, M.~I.\ 2002, 
\apjl, 580, L73 

\bibitem[Johnson(1981)]{johnson81} Johnson, H.~M.\ 1981, \apj, 243, 234 

\bibitem[Johnson(1987)]{johnson87} Johnson, H.~M.\ 1987, \apj, 316, 458 


\bibitem[Kraft et al.(1991)]{kraft91} Kraft, R.~P., Burrows, D.~N., \& Nousek, J.~A.
\ 1991, \apj, 374, 344 

\bibitem[Kundu et al.(1988)]{kundu88} Kundu, M.~R., White, S.~M., Jackson, P.~D., \& Pallavicini, R.\ 1988, \aap, 195, 159 

\bibitem[Mart\'\i n \& Bouy(2002)]{martin02} Mart\'\i n, E.~L., \& Bouy, H. 2002, NewA, 7, 595

\bibitem[McCaughrean et al.(2004)]{mccaughrean04} McCaughrean, M.~J.,  
et al.\ 2004, \aap, 413, 1029 

\bibitem[Meyer \& Meyer-Hofmeister(1999)]{meyer99} Meyer, F., \& Meyer-Hofmeister, E.\ 1999, \aap, 341, L23 

\bibitem[Mohanty \& Basri(2003)]{mohanty03} Mohanty, S., \& Basri, G.\ 2003, \apj, 583, 451 

\bibitem[Mohanty et al.(2002)]{mohanty02} Mohanty, S., Basri, G., Shu, F., Allard, F., \& Chabrier, G.\ 2002, \apj, 571, 469 

\bibitem[Neupert(1968)]{neupert68} Neupert, W.~M.\ 1968, \apjl, 153, L59 

\bibitem[Rutledge et al.(2000)]{rutledge00} Rutledge, R.~E., Basri, G., Mart\'\i n, E.~L., \& Bildsten, L. 2000, ApJ, 538, L141

\bibitem[Sault et al.(1995)]{sault95} Sault, R.~J., Teuben, P.~J., \& Wright, M.~C.~H.\ 1995, Astronomical Society of the
Pacific Conference Series, 77, 433 

\bibitem[Schmitt \& Liefke(2002)]{schmitt02} Schmitt, J.~H.~M.~M., \& Liefke, C.\ 2002, \aap, 382, L9 

\bibitem[Scholz et al.(2003)]{scholz03} Scholz, R.-D., McCaughrean, M.~J., Lodieu, N., \& Kuhlbrodt, B.\ 2003, \aap, 398, L29 

\bibitem[Smith et al.(2001)]{smith01} Smith, R.~K., Brickhouse, N.~S., Liedahl,
D.~A., Raymond, J.~C. 2001, ApJ, 556, L91

\bibitem[Smith et al.(2003)]{smith03} Smith, V.~V., et al.\ 2003, \apjl, 599, L107 

\bibitem[Stelzer(2004)]{stelzer04} Stelzer, B. 2004, \apj, 615, L153

\bibitem[Tagliaferri et al.(1990)]{tagliaferri90} Tagliaferri, G., Giommi, P., \& Doyle, J.~G.\ 1990, \aap, 231, 131 

\bibitem[Volk et al.(2003)]{volk03} Volk, K., Blum, R., Walker, G., \& Puxley, P.\ 2003, \iaucirc, 8188, 2 

\end{thebibliography}
\end{document}